\documentclass{article}

\font\sqi=cmssq8
\def\DR{\rm I\kern-1.45pt\rm R}
\def\DC{\kern2pt {\hbox{\sqi I}}\kern-4.2pt\rm C}
\newcommand{\ben}{\begin{enumerate}}
\newcommand{\een}{\end{enumerate}}
\newcommand{\beq}{\begin{equation}}
\newcommand{\eeq}{\end{equation}}
\newcommand{\bse}{\begin{subequation}}
\newcommand{\ese}{\end{subequation}}
\newcommand{\bea}{\begin{eqnarray}}
\newcommand{\eea}{\end{eqnarray}}
\newcommand{\bc}{\begin{center}}
\newcommand{\ec}{\end{center}}

\newcommand{\bs}{\mbox{\boldmath $\sigma$}}
\newcommand{\bp}{\mbox{\boldmath $\pi$}}
\def\r{r^2_0}

\def\DH{\rm I\kern-1.5pt\rm H\kern-1.5pt\rm I}
\begin{document}
\begin{center}
{\large\bf $3d$ Oscillator and Coulomb systems reduced from K\"ahler
 spaces} \\
\vspace{0.5 cm}
{\large Armen Nersessian$^{1,2}$ and Armen Yeranyan$^1$ }
\end{center}
{\it $^1$ Yerevan State University, Alex  Manoogian St., 1, Yerevan,
375025, Armenia\\
$^2$ Yerevan Physics Institute, Alikhanian Brothers St., 2, Yerevan, 375036,
 Armenia}
\vspace{0.4cm}
\begin{center}
 {\sl\large To the  memory of Professor Valery Ter-Antonyan}
\end{center}
\vspace{0.4cm}
\begin{abstract}
We define the oscillator and Coulomb   systems on
four-dimensional spaces with $U(2)$-invariant K\"ahler metric and
perform their Hamiltonian reduction to the three-dimensional
oscillator and Coulomb systems specified by the presence
of Dirac monopoles. We find the
K\"ahler spaces with conic singularity, where the oscillator and
Coulomb systems on three-dimensional sphere and two-sheet
hyperboloid are originated. Then we construct the
 superintegrable oscillator system on
 three-dimensional sphere and hyperboloid, coupled to monopole,
 and find their four-dimensional origins.
 In the latter
 case the  metric of configuration space  is non-K\"ahler one.
Finally, we extend these results  to the  family of K\"ahler spaces with
conic singularities.
\end{abstract}
\begin{center}
\end{center}
\setcounter{equation}0
\section{Introduction}
The oscillator and Coulomb systems  play a distinguished role in
theoretical and mathematical physics, due to their  overcomplete
symmetry group. The wide number of hidden symmetries provides
these systems  with unique properties, e.g., closed classical
trajectories, the
 degenerated  quantum-mechanical
 energy spectrum, the separability of variables in a few
coordinate systems. The overcomplete symmetry allows to preserve
their exact solvability even after some deformation of the
potential breaking the initial symmetry, or, at least, to simplify
the perturbative calculations. The reduction of these systems  to
low dimensions allows one to construct new integrable systems
with hidden symmetries \cite{perelomov}.

There exist nontrivial generalizations of the oscillator and Coulomb
systems on the
 sphere and the two-sheet hyperboloid (pseudosphere)
\cite{higgs}   given  by the
  potentials
\begin{equation}
V_{Coulomb}=-\frac{\gamma}{r_0}\frac{x_{d+1}}{|{\bf x}|}, \quad
V_{osc}=\frac{\omega^2 \r}{2}\frac{{\bf x}^2}{{x}^2_{d+1}}.
\label{00}\end{equation}
Here ${\bf x}, x_{d+1}$ are the
(pseudo)Euclidean coordinates of the ambient space
$\DR^{d+1}$($\DR^{d.1}$): $\epsilon{\bf x}^2+ x^2_{d+1}=\r$, with
$\epsilon=+1$ for  the sphere, $\epsilon=-1$ for the hyperboloid.

The potential of oscillator has also been generalized for the
complex projective space $\DC P^N$, $N>1$. That is  defined as
follows \cite{bn}
 \beq
V_{osc}=\omega^2 g^{\bar a b}\partial_{\bar a}K \partial_b K ,
\label{op}
\eeq
where $K(z,\bar z)=\log(1+ z \bar z)$ is  K\"ahler potential of $\DC P^N$.
\\
The generalized systems preserves the property
of ``maximal superintegrability"  of the conventional oscillator
and Coulomb systems.
 They have $2d-1$ functionally independent constants of motion (here $D$
 is the dimension of configuration space).
The definition of the oscillator potential (\ref{op}) tells us  to
define the Coulomb potential on $\DC P^N$ as follows
\beq
V_{Coulomb}=-\frac{\gamma}{\sqrt{g^{\bar a b}\partial_{\bar a}K
\partial_b K}}. \label{Cp} \eeq

In some cases one can establish
the nontrivial relation between oscillator and
 Coulomb systems:
the $(p+1)-$dimensional  Coulomb problem can be obtained from the
 $2p-$dimensional oscillator
by the so-called   Levi-Civita (or Bohlin) ($p=1$),
 Kustaanheimo-Stiefel ($p=2)$ and
Hurwitz transformations ($p=4$),
 when  $p=1, 2,4$ \cite{bohlin},  corresponding
  to the reduction
 by the actions of
$Z_2$, $U(1)$ and $SU(2)$ groups, respectively.
To be more precise, these
transformations connect the energy levels of  oscillators with
 the ones  parametric families  of Coulomb-like systems,
specified by the presence of a magnetic flux for $p=1$ \cite{1};
by a Dirac monopole  for $p=2$ (the MIC-Kepler system)\cite{2};
and by a Yang monopole  for $p=4$ \cite{3}(for the review,  see
\cite{ta}).
Among these systems most elegant (and important)
 one is, probably, the MIC-Kepler
system, describing of the relative motion of two Dirac dyons.
It is also relevant
to  the the scattering of two well-separated BPS monopoles and dyons.
The latter system  was considered in a well-known paper  by Gibbons and
Manton \cite{gm}, where the existence of a hidden Coulomb-like
symmetry was established.
Nowdays   the MIC-Kepler system is studied in details as well as
Coulomb one \cite{mic}.

To relate  the four-dimensional oscillator
with  the MIC-Kepler system, we have to perform
its  Hamiltonian reduction  by  the action of  $U(1)$ group
which leads  the canonical symplectic structure on
 $T^*\DC^2$  to the twisted symplectic structure on $T^*\DR^3$
specified by the presence  of  monopole magnetic field
\begin{equation}
\Omega_{can}=dz\wedge d\pi+ d\bar z \wedge d\bar\pi\quad\to
\Omega_{red}= d{\bf x}\wedge d{\bf p} +s\frac {{\bf x} \times
d{\bf x }\times d{\bf  x}}{|{\bf x}|^3}. \label{0}\end{equation}
Here $s$ denotes the value of the generator of the Hamiltonian
action
\begin{equation}
s=\frac{J_0}{2},\quad\quad J_0=i(z\pi-\bar z\bar\pi).
\label{j0}
\end{equation}
The reduced coordinates are
connected with the initial one as follows
\begin{equation}
 {\bf x} =z\bs\bar z,\quad
  {\bf p}=\frac{z\bs\pi +\bar\pi\bs\bar z}{2z\bar z},
\label{ks}\end{equation}
where  $\bs$ denote  Pauli  matrices.\\
Upon this reduction,  the energy surface of the
four-dimensional  oscillator yields
 the one of the  MIC-Kepler system.
Applying  this reduction to the oscillator on
 four-dimensional sphere/hyperboloid  and on
 the complex projective space $\DC P^2$
 we shall get  the MIC-Kepler
system on three-dimensional hyperboloid \cite{np,bn}.

The appearing, in the reduced system,  of the monopole field  is
due to Hamiltonian reduction: that corresponds to the compactification
of the  spatial degree of freedom in the circle, which
generate the magnetic charge. So, the above reduction could be
used for the construction of the three-dimensional
systems with monopole from the four-dimensional systems without it.
Vice versa, on can try to construct the superintegrable
four-dimensional system (without monopole) by the lifting  the given
three-dimensional superintegrable one.

{\it In present paper we analyze the following  question:
 wether exist the
maximally superintegrable systems on four-dimensional
$U(2)-$invariant K\"ahler spaces, whose reductions yield the
(superintegrable) three-dimensional oscillator and Coulomb systems
with monopoles, including the systems on the  configurational
spaces with non-constant curvature.}

For this purpose we reduce the
Hamiltonian system on  the four-dimensional space equipped with
$U(2)-$ invariant K\"ahler metric to the system on
three-dimensional  conformal-flat  space ({\it Section 2}). We
find that the the oscillator and Coulomb systems on the
three-dimensional space, sphere and hyperboloid are originated on
the four-dimensional oscillator and Coulomb systems on some
K\"ahler spaces with conic singularity, so that  (\ref{op}),
(\ref{Cp}) give  us the   well-defined generalizations of the oscillator
and Coulomb  potentials on K\"ahler spaces. However, in the
presence of Dirac monopole  field arose due to Hamiltonian
reduction, the trajectories of the three-dimensional systems
becomes unclosed. Hence, in general, these systems are not
superintegrable ones.
 On the other hand, one can define   the
``maximally superintegrable" generalization of the
three-dimensional oscillator a with Dirac monopoles, which is
originated  in the four-dimensional system with non-K\"ahler
metrics ({\it Section 3}). We also find the family of
superintegrable four-dimensional oscillators,
 which yield the ``maximally
superintegrable"  oscillator with monopoles, on the
three-dimensional spaces with non-constant curvature ({\it Section
4}).

\setcounter{equation}{0}
\section{ $3d$ systems with monopoles
from  of $U(2)-$invariant K\"ahler spaces}
As we mentioned in the
Introduction, the Hamiltonian reduction, by the action of $U(1)$ group,
  of the eight-dimensional
canonical symplectic structure
yields the six-dimensional canonical symplectic structure twisted
by the magnetic field of the Dirac monopole (\ref{0}).
Particular consequence of
this reduction is the reduction of the energy surface of the
 oscillator (on $\DC^2$, $S^4$, $\DC P^2$) to the
energy surface of the MIC-Kepler system (on $\DR^3$ and $AdS_3$).

In this Section we would like to reveal which sort of system arise
upon   $U(1)$ Hamiltonian reduction of  the four-dimensional
mechanical systems on the spaces with $U(2)-$invariant K\"ahler
metrics. Particulary, we hope do find, in this way,
the four-dimensional origins of the three dimensional
oscillator and Coulomb systems on Euclidean spaces,
spheres and hyperboloids (which are superintegrable systems).\\

The K\"ahler potential of the  $U(2)$-invariant K\"ahler spaces
$M$,  $dim_{\DC}M_0=2$ is (in the appropriate local coordinates
$z^a$, $a=1,2$) of  the form $K(z\bar z)$. Hence, the
corresponding metric  reads
 \beq
  g_{a\bar b}=\frac{\partial^2
K(z\bar z)}{\partial z^a{\partial\bar z}^b}= a\delta_{a\bar b}+
a'{\bar z}^a  z^b, \qquad {\rm where }\quad  a=dK'(y)/dy, \quad
a'=a'(y) .
\label{gk}\eeq
The particular cases of these spaces are
the Euclidean space $\DC^2$ (when $K=z\bar z$)
and the complex projective space $\DC P^2$
(when $K=\log(1+z\bar z)$).

 The motion of particle on $M$ in the
$U(2)$-invariant potential field $V$ is described by the
  following Hamiltonian system
\begin{equation}
\Omega_{can}=dz^a\wedge d\pi_a +d{\bar z}^{a}\wedge d{\bar\pi}_{a},\quad {\cal H}=  g^{a \bar b}\pi_a{\bar \pi}_b
+ V(z{\bar z}).
\label{o0}\end{equation}
The Nether constants of motion  corresponding to $U(2)$ symmetry are given by the generators
\beq
{\bf J}=iz{\bs}\pi-i\bar\pi{\bs}\bar z,
\quad J_0=iz\pi-i\bar z\bar\pi\; :\quad
\begin{array}{c}
\{J_0, {J}_k\}=0,\quad \{J_k,J_l \}=2\epsilon_{klm}J_m,\end{array}
\label{nether}\eeq
where $\bs$ denotes  standard Pauli matrices.\\

In order to perform the Hamiltonian reduction of  this  system by
the action of the generator  $J_0$, we have to fix its level
surface,
 \beq
J_0=2s, \eeq and then factorize the level surface of  by  the
action of vector field $\{J_0,\;\}$.
 The resulting six-dimensional
phase space $T^*M^{\rm red}$ could  be parameterized by the following
$U(1)$-invariant
 functions:
\beq
 {\bf y}=z\bs{\bar z},\quad
 {\bp}=\frac{z\bs\pi +\bar\pi\bs\bar z}{2z\bar z}\;
:\quad \{{\bf y}, J_0\}=\{{\bp}, J_0\}=0.
\label{ksc}\end{equation}
In these coordinates
 the reduced  symplectic structure and the generators of the
 angular momentum are given by the expressions
\begin{equation}
\Omega_{\rm red}=d{\bp}\wedge d{\bf y} +
s\frac{{\bf y}\times
 d{\bf y }\times d{\bf y}}{|{\bf y}|^3},
\quad {\bf J}_{red}={\bf J}/2= {\bp}\times{\bf y} + s\frac{{\bf
y}}{ |{\bf y}|  }. \label{ss2}\end{equation}
 The
reduced Hamiltonian is given by the expression
\begin{equation}
 {\cal H}_{red}=\frac{1}{a}\left[{y}{\bp}^2 -
 b{({\bf y} {\bp})^2}\right]
+ s^2\frac{1-b y}{a y}+V(y),
\quad{\rm where}\quad y\equiv |{\bf y}|,\quad
b=\frac{a'(y)}{a+y a'(y)}.
\label{hred}\end{equation}
Hence, the reduced
system is specified by the presence of a Dirac  monopole.

Let us perform the canonical
 transformation $({\bf y},{\bp})\to({\bf{ x}},
{\bf{ p}})$  to the coordinates, where the metric takes a
conformal-flat form:

\beq
{\bf { x}}= f(y){\bf y},\quad {\bp}= {f} {\bf {  p}}
+\frac{df}{dy} \frac{({\bf y{ p}})}{y}{{\bf y }},
 \label{ct}\eeq
where
\beq
\left(1+\frac{yf'(y)}{f}\right)^2= 1+\frac{ y a'(y)}{
a}\quad \Rightarrow\quad \left(\frac{d\log x}{dy}\right)^2=
\frac{d\log ya(y)}{ydy}.
\eeq
Notice, that $x<1$.

 In the new
coordinates the Hamiltonian takes the form
 \beq {\cal H}_{red}=
 \frac{{x}^2(y)}{ya(y)}{\bf{ p}^2}+\frac{s^2}{y(a+y
a'(y))} +V\left(y(x)\right).
\eeq
In order to express the $y$, $a(y)$, $a'(y)$ via ${ x}$ it is
convenient to introduce the function
\beq {\tilde A}(y)\equiv \int
(a+ya'(y))y f(y)dy
 \eeq and consider its Legendre transform ${A}({
x})$, \beq
 { A}({ x})={ x} a(y)y-\tilde A(y)
\eeq
 Then, we get immediately
  \beq \frac{d{ A}({ x})}{d{ x}}=
a(y)y, \quad { x}\frac{d^2{ A}}{d{ x}^2}= {y\sqrt{a(a+y a'(y))}}.
\eeq
By the use of  these expressions, we can present  the reduced
 Hamiltonian system as follows
 \beq
 {\cal H}_{red}=\frac{{x}^2}{N^2}{\bf{ p}}^2 +
\frac{s^2}{\left(2{x}N'(x)\right)^2} + V\left(y(x)\right),\quad
\Omega_{\rm red}=d{\bf p}\wedge d{\bf x} + s\frac{{\bf  x}\times
 d{\bf x }\times d{\bf  x}}{|{\bf x}|^3},
\label{rh}\eeq
 where
\beq N^2({ x})\equiv\frac{d{A}}{d{x}}.
 \eeq
 The K\"ahler potential
of the initial system is connected with $N$ via the  equations
\beq
\frac{dK}{d{x}}=\frac{N^3(x)}{2{x}^2N'(x)},\quad \frac{d\log
y}{d{x}}=\frac{N}{2{x}^2N'(x)}. \label{corresp}\eeq

Let us postulate, that the ``oscillator potential" on the spaces under
consideration  by the same formula, as on the complex projective spaces,
(\ref{op}).
Then, upon the reduction it will reads as follows
\beq V_{osc}=
\omega^2\partial_{\bar a}Kg^{\bar a b}\partial_{b}K=
\omega^2\frac{ya^2(y)}{a(y)+ya'(y)}= \left(\omega\frac{N^2}{2{
x}N'(x)}\right)^2.
 \label{rop} \eeq
Similarly, we could choose the ``Coulomb potential" (\ref{Cp}),
and get its  reduced version
 \beq
 V_{Coulomb}=-\gamma\frac{2xN'(x)}{N^2(x)}. \label{rcp}\eeq

%
%
Further study we will need to consider the
classical trajectories of the (reduced) system, in order to check their
their closeness (closeness of trajectories is the explicit
 indication of superintegrability).

For this purpose it is convenient
 to  direct
 the $x_3$ axis  along ${\bf J}$,
i.e. assume, that $J=J_3$. Upon this choice of coordinate system
one has
 \beq
\frac{ x_3}{ x}=\frac sJ.
 \label{tan}\eeq
 Then, after obvious
manipulations, we get
 \beq \frac{d\phi}{dt}=\frac{2J}{N^2}, \quad
{\cal E}=\frac{J^2-s^2}{N^2} +\frac{J^2}{{ x}^2 N^2}\left(\frac{d{
x}}{d\phi}\right)^2+\frac{s^2}{(2xN'(x))^2}+V( x),\label{teq}
\eeq
 where
\beq
 \phi={\rm arctan} \frac{ x_1}{ x_2}
 \label{phi}
 \eeq
From the expression  (\ref{teq}) we find,
\beq
|\frac{\phi}{J}|=\int\frac{d{ x}}{{x}\sqrt{({\cal
E}-V_{eff})N^2-J^2+s^2}} \qquad {\rm where}\quad V_{eff}=V({ x})+
\frac{s^2}{(2 xdN/d{ x})^2}.
 \label{tg}\eeq

 \subsubsection*{Euclidean space}
Let us consider the simplest case, when the reduced configuration
space is $\DR^3$, i.e. $N=\sqrt{2}x$.
 In that case the reduced Hamiltonian reads
  \beq {\cal H}_{red}=\frac{{\bf p}^2}{2} + \frac{s^2}{8 { x}^2}
  + V\left( x \right). \eeq
The trajectories of the system are defined by the equations
(\ref{tan}) and
 \beq
 |\frac{\phi}{J}|
 = \int\frac{d{
x}}{{x}\sqrt{2({\cal E}-V)x^2-J^2+3s^2/4}}
 \label{tgflat}\eeq
   The K\"ahler potential and
metric of the original four-dimensional system are of the form
\beq
K= (z\bar z)^4,\quad g_{a\bar b}=4(z\bar
z)^2\left[(z\bar z)\delta_{a\bar b}+ {3 \bar z^a z^b}\right].
\label{4kf}\eeq
 Hence, the systems on $\DR^3$ is originated on the
K\"ahler conifold\footnote{We thank Dmitry Fursaev for this
remark.}. \\
Notice, that the oscillator and Coulomb potentials
(\ref{op}), (\ref{Cp}) takes, on this conifold, the following form
\beq
 V_{Coulomb}=-\frac{\gamma}{(z\bar z)^2},\quad
V_{osc}=\omega^2(z\bar z)^4. \eeq
 Upon reduction
they yield the oscillator and Coulomb potentials on $\DR^3$!\\

On the other hand, for the
$$V_{eff}=\frac{s^2}{x^2}+V(x)$$ one has
\beq
|\frac{\phi}{J}|= \int\frac{d{ x}}{{x}\sqrt{2({\cal
E}-V)x^2-J^2}},
 \label{tgflat1}\eeq
 so that the form of trajectory, $\phi(x)$, is independent on
 ``monopole number" $s$.

 Hence, the  well-defined monopole generalization of
 the system on $\DR^3$ with potential $V(x)$ reads
 \beq
 {\cal H}_{s}=\frac{\bf{ p}^2}{2}+
\frac{s^2}{2{x}^2} + V(x),\quad \Omega_{\rm red}=d{\bf p}\wedge
d{\bf x} + s\frac{{\bf  x}\times
 d{\bf x }\times d{\bf  x}}{|{\bf x}|^3}.
\label{micV}\eeq
 Its  four-dimensional origin is formulated  as follows,
 \beq
 \Omega_{can}=dz^a\wedge d\pi_a +d{\bar z}^{a}\wedge d{\bar\pi}_{a},
\qquad {\cal H}=g^{\bar a b}\bar\pi_a\pi_b+\frac{3J_0^2}{16(z\bar
z)^4}+ V(z\bar z), \eeq
where $g_{\bar a b}$ is  given by (\ref{4kf}).\\

  \setcounter{equation}{0}
\section{Sphere and hyperboloid}
In this Section we consider the particular case of our construction,
when the reduced system (\ref{rh})  is formulated  on the
three-dimensional sphere or (two-sheet) hyperboloid.

For this purpose  we  choose
the following value of $N$:
 \beq
  N={ 2{\sqrt 2} r_0  x}/(1+\epsilon {
x}^2),\;\epsilon=1, \ -1  . \label{NS}
\eeq
 Here $\epsilon=1$
corresponds to the sphere, $\epsilon=-1$ corresponds to the
two-sheet hyperboloid.

The corresponding Hamiltonian is of the form
 \beq
{\cal H}_{red}= \frac{(1+\epsilon x^2)^2}{8\r}\left( {{\bf
p}^2} + \frac{s^2}{4{ x}^2}\right)+ V\left( x \right) +\frac{
s^2x^2}{2\r (1-\epsilon x^2)^2}+\frac{\epsilon s^2}{8\r }.
\label{hs}\eeq
Solving the equations (\ref{corresp}) we could find
the K\"ahler space, where the system (\ref{rh}) is originated. It
is defined by the following K\"ahler potential and metric
 \beq
  K=\frac{\epsilon\r }{2}\log(1+ 4 \epsilon(z\bar z)^4),
 \quad g_{a\bar
b}=\frac{8\r (z\bar z)^2}{1+4\epsilon(z\bar z)^4}
\left[(z\bar z)\delta_{a\bar b}+\frac{3-4 \epsilon (z\bar
z)^4}{(1+ 4\epsilon (z\bar z)^4)}\bar z^a z^b\right].
\label{4k}\eeq
 Hence, the systems on  sphere and
two-sheet hyperboloid are also originated on the K\"ahler
conifolds.

On these conifolds the oscillator and Coulomb potentials
(\ref{op}), (\ref{Cp}) read as follows
 \beq V_{Coulomb}=-\frac{
\gamma}{\sqrt{2}r_0 (z\bar z)^2},\quad
V_{osc}=2\omega^2\r {(z\bar z)^4}. \eeq
 Upon reduction
to the sphere and hyperboloid they take the form
 \beq
V_{Coulomb}=-\frac{{\sqrt{2}} \gamma}{r_0}\frac{1-\epsilon
x^2}{2x}, \quad V_{osc}=\omega^2\r\frac{2 x^2}{(1-\epsilon
x^2)^2}.
 \label{csh}\eeq
  These potentials are precisely  Coulomb
and oscillator potentials on the sphere and hyperboloid (\ref{00})
written in conformal-flat coordinates.
 {\it Hence, the
oscillator/Coulomb system on the conifold (\ref{4k}), reduces, for
 $J_0=0$, to the oscillator/Coulomb system
 on three-dimensional sphere/hyperboloid.}
Hence, the initial four-dimensional oscillator and Coulomb systems
are  superintegrable ones,
when the constant of motion $J_0$ takes the value $J_0=0$.\\

When $J_0\neq 0$, the relation  between
three- and four- dimensional systems
is  more complicated, and
needs separate consideration of the oscillator and Coulomb cases.\\

Let us consider, at first, the case of  oscillator.
 For a checking the superintegrability, let us
clarify, wether  trajectories of the reduced system are closed.

Substituting (\ref{NS}) and (\ref{csh}) in  (\ref{tg}), we get
\beq |\frac{\phi}{J}|=\int
\frac{du}{\sqrt{-4\r (
\omega^2\r +2{\cal
E})+(8\r {\cal E}+l^2)u- (s^2+l^2)u^2}} \label{ot}\eeq
where
\beq l^2=4(J^2-s^2),\quad 4u=({ x}+1/{ x})^2. \label{od}\eeq
From this expression we easily get
\beq \left( x+\frac{1}{
x}\right)^2=8\frac{2\r {\cal E} + J^2-s^2}{4J^2-3s^2}
\left(1+\sqrt{1-4\frac{(2\r {\cal
E}+r^4_0\omega^2)(4J^2-3s^2)}{(2\r{\cal
E}+J^2-s^2)^2} }\sin 2\sqrt{1-\frac{3s^2}{4J^2}}\;|\phi |
\right) \eeq
 Hence, trajectories are  closed only when
$$
\sqrt{1-\frac{3s^2}{4J^2}} \quad {\rm is }\quad {\rm
rational}\quad{number}.
$$
Particulary, trajectories are closed in in the ``ground state'',
i.e. for $s=J$. In this case they  belong to ``equatorial plane'',
${ x}_3 ={ x}$. Hence,  the Hamiltonian system (\ref{rh}) on
sphere/hyperboloid with the  oscillator potential  is not
superintegrable for the arbitrary value of monopole number $s$.

However, one can get  the monopole generalization of oscillator
whose trajectories are closed for any $s$,
 choosing the potential
\beq
V^s_{osc}=V_{osc}+\frac{3s^2}{4{ x}^2(dN/d{ x})^2}. \eeq
In that
case the trajectories  are given by the expression
\beq
\left(x+\frac{1}{ x}\right)^2=2\frac{2\r{\cal E} + J^2-s^2}{2J^2}
\left(1+\sqrt{1- 16 J^2\r \frac{2{\cal E}+\omega^2\r}
{(2\r {\cal E}+J^2-s^2)^2} }\sin\; 2|\phi| \;
\right), \eeq
i.e. they are closed for any $s$.

Hence, the superintegrable   generalization of the Higgs
oscillator specified
 by the presence of Dirac monopole
  is defined by the Hamiltonian
\beq
{\cal H}^{\epsilon}_{MIC-osc}=\frac{(1+\epsilon
x^2)^2}{8\r} \left( {\bf{p}^2}+ \frac{s^2}{x^2}\right)
+(\omega^2\r + \frac{s^2}{4\r}  )\frac{2x^2}{(1-\epsilon
x^2)^2},\qquad \epsilon=\pm 1,
 \label{mice}\eeq
where $\epsilon=1$
corresponds to the sphere and $\epsilon=-1$ to the hyperboloid.

It is originated in the Hamiltonian given by the  expression
\beq
{\cal H}=g^{\bar a b}\bar\pi_a\pi_b+\frac{3J_0^2}{16R(z\bar z)}+
\omega^2{K_ag^{a\bar b}K_{\bar b}},\quad R=
\frac{32\r (z\bar z)^4}{(1+4\epsilon (z\bar z)^4)^2}\;.
 \label{oimp}\eeq
 There is an important difference  of the above reduced oscillator
  from the one on  the
 $\DR^3$.

Namely, the four-dimensional  system with ``frequency'' $\omega$
yields the three-dimensional oscillator with ``frequency" dependent on
 the ``monopole number" $s$
$$
\omega_s= {\sqrt{\omega^2 + \frac{s^2}{4r^4_0}}},
$$
while the frequency of the oscillator reduced to $\DR^3$
 is independent on $s$.\\

Now, let us  consider the   system  with
Coulomb potential (\ref{Cp}) on the conifold with K\"ahler
structure (\ref{4k}).
 After Hamiltonian reduction it yields the
three-dimensional system  with Hamiltonian (\ref{hs}) where
$V=V_{Coulomb}$ (\ref{csh}).

On the level surface $s=0$ (i.e. in the absence of Dirac monopole),
 the reduced system coincides with the standard Coulomb system on the
sphere/hyperboloid. Therefore it is superintegrable one.
On the level surface  $s\neq 0$ (i.e. in the presence of monopole)
the potential of the reduced system is the
superposition of the Coulomb potential and
of the oscillator one, proportional to
$s^2/r^4_0$! So, it is not surprising,
 that  the expression
 for trajectories of the reduced system, $\phi=\phi(r)$
 is given by elliptic integral...

  On the other hand, there are the
superintegrable MIC-Kepler systems on the sphere and hyperboloid,
given by the Hamiltonian
 \beq {\cal H}^{\epsilon}_{MIC}=\frac{(1+\epsilon
x^2)^2}{8\r} \left( {\bf{p}^2}+ \frac{s^2}{x^2}\right) -\frac{
\gamma}{r_0}\frac{1-\epsilon x^2}{2x},
\quad\epsilon=\pm 1, \label{miceC}\eeq
 where $\epsilon=1$
corresponds to the sphere \cite{kur} and $\epsilon=-1$ to the
hyperboloid \cite{np}.

To recover this system,  we can try to
 modify the initial Coulomb system, transiting to
the non-K\"ahler metric, as in the case of oscillator
(compare with  (\ref{oimp})),
\beq
{\cal H}=g^{\bar a b}\bar\pi_a\pi_b+\frac{3J_0^2}{16R(z\bar z)}+
\frac{\gamma}{\sqrt{K_ag^{a\bar b}K_{\bar b}}}, \qquad
R=\frac{32\r (z\bar z)^4}{(1+\epsilon (z\bar z)^4)^2}.
\label{MICK} \eeq
However, in that case  the reduced Hamiltonian is given by
the one of  MIC-Kepler system (\ref{mice}) with additional
oscillator potential.
So even modified Coulomb system is not exactly solvable for any value of
$J_0$.\\

We have found the four-dimensional oscillator and Coulomb systems on
appropriate K\"ahler conifolds,
which result, after Hamiltonian reduction, in the
 oscillator and Coulomb systems
 on  three-dimensional sphere
 and two-sheet hyperboloid.
The appearance, in the reduced system, of the Dirac monopole,
 breaks the superintegrability of the system.
 However, the superintegrability of the oscillator system, in
 opposite to the  Coulomb one,
 could be restored by the transition to the
non-K\"ahler metric.

\setcounter{equation}{0}
\section{Family of oscillator systems}
The results of previous Section could be easily extended to the
K\"ahler space whose metric is defined by the
 potential
\beq
K=\frac{\epsilon\r}{2}\log (1
+4\epsilon(z\bar z)^n), \quad \epsilon=\pm 1,\quad n>0.
\label{la}\eeq
 For
$n=1$ the potential (\ref{la})
  defines the Fubini-Studi metric of the two-dimensional
complex projective space  $\DC P^2$ (for $\epsilon=1$) and its
noncompact version, the four-dimensional Lobacewski space ${\cal
L}_2$ (for $\epsilon=-1$).
 These spaces are  of the constant curvature ones,   and have the
 the isometry group $SU(3)$ for $\epsilon=1$ and $SU(1.2)$ for
$\epsilon=-1$.\\
 The case  $n=4$ was considered in previous Section.
  The system on such spaces results, after Hamiltonian reduction,
  in the ones on sphere($\epsilon=1$)
 or two-sheet hyperboloid $\epsilon =-1$
 (which have the constant curvature, and the isometry group
$SO(4)$ and $SO(1.3)$ respectively).
For any other $n$ both initial and reduced spaces have
non-constant curvature
and conic singularity.

The Hamiltonian systems on the spaces with K\"ahler potential
(\ref{la}) results, after reduction,
 in the the three-dimensional  systems (\ref{rh}), with
\beq
N^2=2n\r\frac{x^{\sqrt{n}}}{(1+\epsilon x^{\sqrt n})^2}.
\label{Nc}\eeq
Hence, the metric of the reduced configuration space is given
by the expression
 \beq
ds^2=\frac{2n\r x^{\sqrt{n}-2}(d{\bf x})^2}{(1+\epsilon
x^{\sqrt{n}})^2}, \eeq
so that for  $n\neq 4$ it  has a conifold structure
\footnote{In the vicinity of
singularity this metric could be presented  as follows
$ds^2=dR^2+R^2d\bar{\Omega}^2$, where $R=r^{\sqrt{n}/2}
/{\sqrt{n}/2}$, $d\bar{\Omega}^2=(n/2)^2 d\Omega^2$ where
$d\Omega^2$ is a metric on $S^2$. Hence, the solid angle around
singularity is equal to $n\pi$, instead of $4\pi$ (D.Fursaev).}.

The  oscillator potential
(\ref{rop}) looks as follows:
 \beq
 V_{osc}={2\r\omega^2} (z\bar z)^n.
\eeq
It is reduces to the following form
 \beq
V^{red}_{osc}={2\r \omega^2}\frac{x^{\sqrt{n}}}{(1-\epsilon
x^{\sqrt{n}})^2} .
\eeq

The  trajectories of the reduced oscillator     are  given by the
 expression
  \beq |\frac{\phi}{J}|= \int
\frac{du}{\sqrt{-n\r (\r\omega^2+
2 {\cal E})
+(2n\r{\cal E}+l^2)u- (\frac{4}{n}s^2+l^2)u^2}},
\eeq
 where
 \beq
l^2=4(J^2-s^2),\quad 4u=({x}^{\sqrt{n}/2} +1/{ x}^{\sqrt{n}/2})^2.
\label{odf}\eeq
 From this expression we easily get
  \beq
\left(x^{\sqrt{n}/2}+{ x^{-\sqrt{n}/2}}\right)^2=
\nonumber\eeq
\beq\frac{n\r{\cal
E}+ 2(J^2-s^2)}{2(J^2-s^2(1-1/n))}
\left(1+\sqrt{1-4n\r\frac{(2{\cal
E}+{\r\omega^2}) (J^2-s^2(1-1/n))}{({\cal
E}\r+(J^2-s^2))^2} }\sin2\sqrt{1-\frac{(n -1)s^2}{n
J^2}}\; |\phi| \right).
 \eeq
Hence, trajectories are  closed when the
following condition holds
$$
\sqrt{1-\frac{(n-1)s^2}{nJ^2}}
 \quad {\rm is }\quad {\rm rational}\quad{number}.
$$

Therefore, trajectories are closed for any $s$ only when  $n=1$,
i.e. on  the  complex projective space $\DC P^2$ (for $\epsilon=1$)
and on  its noncompact version,
 four-dimensional Lobachewski
 space ${\cal L}^2= SU(1.2)/U(1)\times SU(2)$.
In this case  the potential  takes quite simple form,
$V=2\omega^2\r z\bar z$.
 The closeness of trajectories are
due to the hidden symmetries of the system, given by the
expressions \cite{bn}
\beq
{\bf I}=\frac{J_+{\bs}J_-}{2\r} +
2\r\omega^2 z{\bs}\bar z \;\quad J^+_a=\pi_a+\epsilon(\bar
\pi\bar z)\bar z^a,\quad J^-=\bar J^{+}, \label{micalg}\eeq
were $J^{\pm}_a$ are the translation generators.
The reduced
Hamiltonian is of the form
\beq
{\cal H}_{\rm red}=
\frac{x(1+\epsilon x)^2{\bf{ p}^2}}{2\r}+
s^2\frac{(1+\epsilon x)^4}{2\r x(1-\epsilon x)^2} +
\frac{2\r\omega^2 x}{(1-\epsilon x)^2}.
\label{rocpn}\eeq
Fixing the energy surface ${\cal H}=E_{\rm osc}$ of the reduced
system, we can transform it in the
MIC-Kepler system on hyperboloid, given by the
Hamiltonian (\ref{mice}).\\

For the $n\neq 1$, we could  get the superintegrable oscillator with
monopole,  choosing the the potential
 \beq
V^s_{osc}=V_{osc}+\frac{(n -1)}{n}\frac{s^2}{{ x}^2(dN/d{ x})^2}.
\eeq
In that case the Hamiltonian of the reduced system reads
 \beq
 {\cal H}=
 \frac{(1+\epsilon x^{\sqrt{n}})^2}{2n\r x^{\sqrt{n}-2}}
{{\bf p}^2} +
s^2\frac{(1 +\epsilon x^{\sqrt{n}})^4}{2n\r x^{\sqrt{n}}
 (1-\epsilon x^{\sqrt{n}})^2}+{2\r\omega^2
}\frac{x^{\sqrt{n}}}{(1-\epsilon x^{\sqrt{n}})^2} ,
\eeq
while the trajectories are  given  by the equation
\beq
(x^{\sqrt{n}/2}+x^{-{\sqrt{n}/2}})^2
=\frac{n\r {\cal E} +
2(J^2-s^2)}{J^2} \left(1+\sqrt{1- 4n\r\frac{(2{\cal
E}+{\r \omega^2)J^2}}
{(n\r {\cal E}+
2(J^2-s^2)^2} }\sin\; 2|\phi| \; \right).
\eeq
 This  superintegrable oscillator with monopole is  originated in
the four-dimensional system with Hamiltonian
\beq {\cal H}=g^{\bar
a b}\bar\pi_a\pi_b+\frac{(n-1)J_0^2}{4n R(z\bar z)}+
{2\r\omega^2} (z\bar z)^n, \qquad R=
\frac{2n^2\r(z\bar z)^4}{(1+4\epsilon (z\bar z)^n)^2}
\label{MICKn} \eeq
where $g^{\bar a b}$ is defined by the K\"ahler
potential (\ref{la}).
\vspace{5mm}
\section{Conclusion}
 We considered the reduction of the mechanical systems on
four-dimensional K\"ahler spaces with $U(2)$ isometry to the
three-dimensional systems, paying special attention to the
``oscillator" and ``Coulomb" systems, defining their   potentials by
the expressions (\ref{op}) and (\ref{Cp}), respectively.
 From the
previous study \cite{bn} it was known, that such a ``oscillator"
potential defines the well-defined
 superintegrable system on $\DC
P^n$, and is distinguished with respect to supersymmetrization as well.
Since the  Hamiltonian reduction by the action of $U(1)$ group
generates, in the resulting three-dimensional system, the magnetic field
of Dirac monopole. We hoped to
find, in this way, the  superintegrable generalizations of
oscillator and Coulomb systems on curved spaces, specified by the
presence of Dirac monopole.
Particularly, we found the four-dimensional K\"ahler spaces (with
 conic sigularities)
where the three-dimensional oscillator and Coulomb systems on the $\DR^3$,
$S^3$, $\DH^3$ are originated, and established, that the original
oscillator and Coulomb potential are, indeed, given by
(\ref{op}) and (\ref{Cp}).
However, when these, four-dimensional systems results in the
three-dimensional ones specified by the presence of Dirac monopoles,
their trajectories become unclosed. In other words,  the monopole field
breaks superintegrability of the system.
 On the other hand, in the case of oscillator,  transiting from the
 K\"ahler  metric to the appropriate non-K\"ahler one, we can restore
  superintegrability of systems (both initial and reduced one),
  but we can't do the same for the Coulomb system.
We also extended these consideration for the some parametric
family of K\"aher spaces  including previous ones as a particular
case. We found that the unique representative of this family,
where the oscillator is superintegrable, is the complex projective space
$\DC P^2$ (and its non-compact version, Lobachewski space ${\cal L}_2
=SU(2.1)/SU(2)\times U(1)$. The energy surface
 of the oscillator on this four-dimensional space leads to the
  energy surface of
 the MIC-Kepler system on three-dimensional hyperboloid.
Let  us notice, that while previous  superintegrable generalizations of
oscillator systems were formulated on constant curvature spaces,
 the  superintegrable oscillators  constructed in the
present paper,  could  have configuration spaces with  nonconstant
curvature and conic singularities.

\subsection*{{ Acknowledgments}}
We thank  Dmitry Fursaev for the useful conversations, and Levon Mardoyan
for the interest in work. The work of
A.N is partially  supported by grants INTAS 00-00262 and ANSEF PS81.

\end{document}